\documentclass[10pt]{article}
\usepackage[T1]{fontenc}
\usepackage[utf8]{inputenc}
\usepackage[english]{babel}
\usepackage{array}
\usepackage{float}
\usepackage{url}
\usepackage{multirow}
\usepackage{amsmath}
\usepackage{amsthm}
\usepackage{amssymb}
\usepackage{geometry}
\usepackage{setspace}
\usepackage[unicode=true,
 bookmarks=true,bookmarksnumbered=false,bookmarksopen=false,
 breaklinks=false,pdfborder={0 0 0},pdfborderstyle={},backref=false,colorlinks=false]
 {hyperref}

\makeatletter

\let\SF@@footnote\footnote
\def\footnote{\ifx\protect\@typeset@protect
    \expandafter\SF@@footnote
  \else
    \expandafter\SF@gobble@opt
  \fi
}
\expandafter\def\csname SF@gobble@opt \endcsname{\@ifnextchar[
  \SF@gobble@twobracket
  \@gobble
}
\edef\SF@gobble@opt{\noexpand\protect
  \expandafter\noexpand\csname SF@gobble@opt \endcsname}
\def\SF@gobble@twobracket[#1]#2{}
\providecommand{\tabularnewline}{\\}

\usepackage{url}
\theoremstyle{plain}

\usepackage{lscape}
\usepackage[bottom]{footmisc}
\usepackage{multirow}
\usepackage{array}

\newtheorem{assumption}{Assumption}

\newtheorem{ass}{Assumption}

\makeatother

\begin{document}
\title{Lagrange multipliers in Maximum likelihood estimations and Least squares problems with Constraints\thanks{An earlier version of this paper was presented at 2026 Kansai econometrics conference, Spring Meeting of the Operations Research Society of Japan, and Spring Meeting of the Japanese Statistical Association. I am grateful to the participants for their valuable comments and suggestions.}}
\author{Takeshi Fukasawa\thanks{Waseda Institute for Advanced Study, Waseda University. 1-21-1, Nishiwaseda, Shinjuku, Tokyo, Japan. E-mail: fukasawa3431@gmail.com.\protect \\
This study was supported by JSPS KAKENHI Grant Number JP24K22629.}}
\maketitle
\begin{abstract}
This study investigates a statistical property of Lagrange multipliers in constrained Maximum Likelihood Estimation (MLE) and Least Squares (LS) problems from the perspective of numerical optimization. Building on large-sample theory, we show that the associated Lagrange multipliers converge to zero as the sample size increases, provided the distribution is correctly specified in MLE or the residuals are normally distributed in LS. Although this asymptotic behavior has long been recognized in statistics, it has received little explicit attention in numerical optimization and has rarely been exploited in algorithmic design. Importantly, the insight extends beyond classical low-dimensional settings: even in modern high-dimensional applications---such as deep learning---where the number of parameters may exceed the sample size, the same reasoning applies provided the generalization performance is good.

This observation has two main implications. First, many constrained optimization algorithms, including the Augmented Lagrangian Method, Sequential Quadratic Programming, and Interior Point methods, require initial values for the multipliers, and choosing zero is statistically justified. Numerical experiments for constrained regressions and dynamic discrete choice model estimations support this implication by showing that initializing multipliers at zero usually lead to stable and efficient performance. Second, penalty-based approaches that convert constrained problems into unconstrained ones can perform well when the true multipliers are small. This helps explain why penalty-based methods often perform well in practice.


\pagebreak{}
\end{abstract}

\section{Introduction}

Maximum Likelihood Estimation (MLE) and Least Squares (LS) problems with constraints arise in a wide range of fields. Examples include constrained regressions (linear/logistic regressions) in machine learning and statistics (e.g., \cite{geyer1991constrained,na2025statistical,shapiro2000asymptotics}), inverse problems in natural science (PDE-constrained optimization; e.g., \cite{de2015numerical}), and structural estimations of economic models in economics (e.g., \cite{rust1987optimal,su2012constrained}). In recent years, deep learning techniques have also been applied to these problems (e.g., Physics-Informed Neural Networks (PINNs); e.g., \cite{raissi2019physics,lu2021physics}). Motivated by these developments, recent studies have explored increasingly sophisticated numerical algorithms for solving constrained optimization problems.

In constrained optimization, the values of the Lagrange multipliers play a crucial role in the performance of numerical algorithms. The current study demonstrates that we can utilize the property that the values of Lagrange multipliers are close to 0 in large samples of the dataset, provided the distribution is correctly specified (for MLE) or the residuals follow a normal distribution (for LS). The property can be easily derived by relying on the large sample theory in statistics. Note that, even in deep learning applications, where the number of parameters can be larger than the sample size, we can obtain analogous results provided that the generalization performance is good.

The property has several practical implications. First, many constrained optimization algorithms, including the Augmented Lagrangian Method (ALM), Sequential Quadratic Programming (SQP), and Interior Point (IP) method, and recently proposed methods for solving stochastic optimization problem (e.g., \cite{na2023inequality,na2025statistical}) require specifying initial candidate values for the Lagrange multipliers, and choosing 0 as the initial values is reasonable. The current study presents numerical results supporting the discussions. Second, penalty-based approaches, which convert constrained problems into unconstrained ones by adding penalty terms (e.g., PINNs with soft constraints for inverse problems), are widely used in applications. Our results show that when the true Lagrange multipliers are small, the solution to the penalized problem can be close to the exact constrained solution. This helps explain why penalty-based methods often perform well in practice (cf. \cite{toscano2025pinns} for the survey of the recent studies on PINNs.).

\subsubsection*{Related literature}

This study is closely connected to recent work on numerical algorithms for constrained stochastic optimization (e.g., \cite{berahas2021sequential,berahas2024stochastic,fang2024fully,na2025statistical}). Unlike these studies, which consider general constrained stochastic optimization, we focus specifically on MLE and LS problems. This narrower focus allows us to exploit the statistical property that Lagrange multipliers tend to be small under the conditions previously mentioned. The insights obtained here may contribute to the development of more efficient numerical algorithms tailored to these widely used problem classes.

The asymptotic behavior of Lagrange multipliers---specifically, their convergence to zero as the sample size increases---was originally established by \cite{aitchison1958maximum} in the MLE setting. However, subsequent research primarily examined this property in the context of statistical hypothesis testing, and no previous studies explicitly discussed its implications from the viewpoint of the numerical optimization, to my knowledge. Moreover, prior work typically assumed a classical statistical setting in which the number of observations far exceeds the number of parameters. In contrast, the present study discusses the implications of this property even in high-dimensional settings where the number of parameters may exceed the sample size, a situation common in modern deep learning applications.

Although the present study focuses on how the statistical properties of the Lagrange multipliers inform widely used existing methods, this property is also useful for developing new algorithms for constrained optimization. Building on earlier work in economics, \cite{fukasawa2026sequential} proposes the Sequential Linearly Constrained (SLC) algorithm, which exploits this property, and showed that the SLC algorithm can be several times faster than alternative approaches in several structural estimation applications in economics.\footnote{The key idea is to solve a linearly constrained optimization problem at each iteration. The advantages of the SLC algorithm are twofold. First, it achieves quadratic local convergence in large samples. Second, it can be implemented without explicitly computing the Jacobian of the constraint function---which may be high-dimensional---by relying instead on numerical directional derivatives and Krylov subspace methods. Note that a related algorithm was mentioned in \cite{murtagh1982projected}, although its convergence properties were not established.} The study also shows that previously proposed methods in economics, including NPL (\cite{aguirregabiria2002swapping}) and EPL (\cite{dearing2025efficient}), implicitly utilize the property.

\medskip{}

The rest of this paper is organized as follows. Section \ref{sec:Problem-setting} derives the statistical properties of the Lagrange multipliers, and Section \ref{sec:Implications} discusses their implications for numerical optimization. Section \ref{sec:Numerical-Experiments} presents the results of numerical experiments, and Section \ref{sec:Conclusions} concludes. Appendix \ref{sec:Appendix} provides additional material, including further details and supplementary results for the numerical experiments.

\section{Constrained optimization problem and Lagrange multipliers\label{sec:Problem-setting}}

We consider the following constrained optimization problem:

\begin{eqnarray}
 & \min_{\theta} & Q\left(\theta\right)=\frac{1}{N}\sum_{i=1}^{N}q\left(w_{i};\theta\right)\label{eq:constrained_opt_problem}\\
 & s.t. & g_{E}\left(\theta\right)=0\nonumber \\
 &  & g_{I}\left(\theta\right)\leq0\nonumber 
\end{eqnarray}

Let $\widehat{\theta}$ denote the solution, and let $g\left(\theta\right)\equiv\left(\begin{array}{c}
g_{E}\left(\theta\right)\\
g_{I}\left(\theta\right)
\end{array}\right)$. Here, we let $g\left(\theta\right)\equiv\left(g_{i}\left(\theta\right)\right)_{i=1,\cdots,n_{E}+n_{I}}$ where $n_{E}$ and $n_{I}$ denote the number of parameters, and let $\mathcal{A}\left(\widehat{\theta}\right)$ denote the set of active constraints. Let $\widehat{\lambda}$ denote the Lagrange multiplier of the problem, and let $\mathcal{L}\left(\theta,\lambda\right)\equiv Q\left(\theta\right)+\lambda^{T}g\left(\theta\right)$ denote the Lagrangian function. Note that the stochastic optimization problem $\min_{\theta}Q\left(\theta\right)=E_{w}\left[q\left(w;\theta\right)\right]\ s.t.\ g_{E}\left(\theta\right)=0,g_{I}\left(\theta\right)\leq0$ can be interpreted as a case where $N$ is infinite.

Here, we assume twice continuous differentiability of $Q$ and $g$, and we impose the following two standard assumptions concerning the constrained optimization problem:

\begin{assumption}[Linear independence constraint qualification]

$\left\{ \nabla_{\theta}g_{i}\left(\widehat{\theta}\right)\right\} _{i\in\mathcal{A}(\widehat{\theta})}$ is linearly independent. 

\label{as:LICQ}
\end{assumption}

\begin{assumption}[Second-order sufficient condition]

$d^{T}\nabla_{\theta\theta^{\prime}}\mathcal{L}\left(\theta,\lambda\right)d>0$ holds for all $d\neq0$ such that $A\left(\theta\right)d=0$, where $A\left(\theta\right)$ denotes a matrix $\left(\nabla_{\theta}g_{i}\left(\widehat{\theta}\right)\right)_{i\in\mathcal{A}\left(\widehat{\theta}\right)}$.

\label{as:SOC}
\end{assumption}

The KKT conditions for the problem are:

\begin{eqnarray}
 &  & \nabla_{\theta}Q\left(\widehat{\theta}\right)+\left(\widehat{\lambda}\right)^{T}\left(\nabla_{\theta}g\left(\widehat{\theta}\right)\right)=0\label{eq:FOC}\\
 &  & g_{i}\left(\widehat{\theta}\right)\leq0\ i=1,\cdots,p\\
 &  & \widehat{\lambda_{\mathcal{A}}}\geq0\\
 &  & \widehat{\lambda_{-\mathcal{A}}}=0
\end{eqnarray}

Let $\theta\equiv\left(\theta_{1},\theta_{2}\right)$, where $\theta_{2}\in\mathbb{R}^{n_{A}}$. Note that the divisions of $\theta$ into $\theta_{1}$ and $\theta_{2}$ are arbitrary. $n_{A}$ denotes the number of active constraints. Then, by (\ref{eq:FOC}), 
\begin{eqnarray}
\nabla_{\theta_{2}}Q\left(\widehat{\theta}\right)+\left(\widehat{\lambda_{\mathcal{A}}}\right)^{T}\left(\nabla_{\theta_{2}}g_{\mathcal{A}}\left(\widehat{\theta}\right)\right) & = & 0\label{eq:FOC_2}
\end{eqnarray}
holds. Under Assumption \ref{as:LICQ}, $\left\{ \nabla_{\theta_{2}}g_{i}\left(\widehat{\theta}\right)\right\} _{i\in\mathcal{A}(\widehat{\theta})}$ are linearly independent. Then, $\nabla_{\theta_{2}}g_{\mathcal{A}}\left(\widehat{\theta}\right)$ is nonsingular, and (\ref{eq:FOC_2}) implies:

\begin{equation}
\left(\widehat{\lambda_{\mathcal{A}}}\right)^{T}=-\left(\nabla_{\theta_{2}}Q\left(\widehat{\theta}\right)\right)\left(\left(\nabla_{\theta_{2}}g_{\mathcal{A}}\left(\widehat{\theta}\right)\right)\right)^{-1}\label{eq:lambda_eq}
\end{equation}

\medskip{}

In the rest of the current paper, we focus on two specific problems: 
\begin{enumerate}
\item Maximum Likelihood Estimation (MLE) $\min_{\theta}Q\left(\theta\right)=-\frac{1}{N}\sum_{i=1}^{N}\ln f\left(w_{i};\theta\right)\ s.t.\ g_{E}\left(\theta\right)=0,g_{I}\left(\theta\right)\leq0$

$f\left(w;\theta\right)$ represents the distribution of the observation $w$ using parameters $\theta$ specified by an analyst. $\ln f\left(w|\theta\right)$ corresponds to $q\left(w;\theta\right)$.
\item Least Squares (LS) problem $\min_{\theta}Q\left(\theta\right)=\frac{1}{2N}\sum_{i=1}^{N}\left(r\left(w_{i},\theta\right)\right)^{2}\ s.t.\ g_{E}\left(\theta\right)=0,g_{I}\left(\theta\right)\leq0$

Here, $r\left(w_{i},\theta\right)\equiv y_{i}-\psi\left(x_{i};\theta\right)$ denotes the residual of the observation $i$, where $x_{i}$ and $y_{i}$ are variables in observation $i$ included in $w_{i}\equiv\left(x_{i},y_{i}\right)$. $\frac{1}{2}\left(r\left(w,\theta\right)\right)^{2}$ corresponds to $q\left(w;\theta\right)$.
\end{enumerate}

\subsection{Values of Lagrange multipliers}

We can show that the Lagrange multipliers take values close to 0 under moderate conditions in both the MLE and LS settings. This section clarifies this point. In both cases, define $\theta^{*}\equiv\arg\max_{\theta}E_{w}\left[q\left(w_{i};\theta\right)\right]\ s.t.\ g_{E}\left(\theta\right)=0,g_{I}\left(\theta\right)\leq0$. $\theta^{*}$ can be interpreted as a solution of the constrained optimization problem when an infinite number of observations are available. Here, we assume that $\widehat{\theta}-\theta^{*}=O_{p}\left(N^{-\frac{1}{2}}\right)$ holds (cf. \cite{shapiro2014lectures} for the $\sqrt{N}$-consistency of the constrained M-estimator.).

Note that, in general, $\nabla_{\theta}Q\left(\widehat{\theta}\right)=0$ must hold for an objective function $Q$ satisfying $Q\left(\theta\right)\geq0$ and $Q\left(\widehat{\theta}\right)=0$, because $\widehat{\theta}$ can be regarded as a local minimizer of $Q$ as long as $g_{E}\left(\widehat{\theta}\right)=0,g_{I}\left(\widehat{\theta}\right)\leq0$. This implies $\widehat{\lambda}=0$ by (\ref{eq:lambda_eq}). The discussion below suggests that, under suitable conditions and in large samples, the original objective functions of the MLE and LS can essentially be reformulated in this manner.

\subsubsection{Maximum Likelihood Estimation (MLE)\label{subsec:MLE}}

Let $f^{*}\left(w\right)$ be the true distribution of the observation $w$. When the model is correctly specified, $\theta^{*}$ satisfies $f\left(\cdot,\theta^{*}\right)=f^{*}\left(\cdot\right)$. Note that analogous discussions hold, even when $w_{i}$ can be divided into $x_{i}$ and $y_{i}$ and we alternatively consider the conditional likelihood $f\left(y_{i};\theta,x_{i}\right)$, (i.e., $\min_{\theta}-\frac{1}{N}\sum_{i=1}^{N}\ln f\left(y_{i};\theta,x_{i}\right)\ s.t.\ g_{E}\left(\theta\right)=0,g_{I}\left(\theta\right)\leq0$) which usually appear in applications.

\subsubsection*{Intuition}

We can intuitively understand why the the Lagrange multiplier is close to 0 in large samples for the MLE. Here, we consider a setting where the number of samples is infinite. In the setting, as is well known, the MLE is equivalent to finding parameters minimizing the Kullback Leibler Information Criterion (KLIC). Hence, we let $Q\left(\theta\right)=KLIC\left(f^{*};f\left(\cdot;\theta\right)\right)\equiv\int\left[\ln\frac{f^{*}\left(w\right)}{f\left(w|\theta\right)}f^{*}\left(w\right)\right]dw$, and consider the constrained optimization problem $\min_{\theta}Q\left(\theta\right)=KLIC\left(f^{*};f\left(\cdot;\theta\right)\right)\ s.t.\ g_{E}\left(\theta\right)=0,g_{I}\left(\theta\right)\leq0$. By the property of KLIC, $Q\left(\theta\right)=KLIC\left(f^{*};f\left(\cdot;\theta\right)\right)\geq0$ holds for all $\theta$. If the distribution $f\left(w;\theta\right)$ is correctly specified, $Q\left(\widehat{\theta}\right)=KLIC\left(f^{*};f\left(\cdot;\widehat{\theta}\right)\right)=0$ additionally holds. Then, the discussions at the beginning of the current subsection suggests $\nabla_{\theta}Q\left(\widehat{\theta}\right)=0$, and consequently $\widehat{\lambda}=0$. Therefore, the Lagrange multipliers should be equal to zero in the infinite-sample limit under the correct specification.

It is worth noting that, although $KLIC\left(f^{*};f\left(\cdot;\widehat{\theta}\right)\right)=0$ can hold, it does not necessarily imply the redundancy of the constraints. If the constraints are not imposed, $\theta$ satisfying $KLIC\left(f^{*};f\left(\cdot;\theta\right)\right)=0$ may not be unique.

\subsubsection*{Formal discussion}

The following formal discussions largely simplify the discussions in \cite{aitchison1958maximum}, while considering the possibility of misspecified distribution of $f$. Because the primary focus of the current study is to clarify $\widehat{\lambda}=O_{p}\left(N^{-\frac{1}{2}}\right)$ and the conditions required to derive the result, in contrast to \cite{aitchison1958maximum} deriving detailed asymptotic distributions of $\left(\widehat{\theta},\widehat{\lambda}\right)$, we take a different approach. Here, we assume finite $Var\left(\nabla_{\theta}\ln f\left(w;\theta^{*}\right)\right)$ and the boundedness of $\nabla_{\theta\theta^{\prime}}\ln f\left(w;\theta\right)$.

First, by differentiating the both sides of $1=\int f\left(w;\theta\right)dw\ \forall\theta$ with regard to $\theta$ at $\theta=\theta^{*}$, we have 
\[
0=\int\left(\nabla_{\theta}f\left(w;\theta^{*}\right)\right)dw=\int\left(\nabla_{\theta}\ln f\left(w;\theta^{*}\right)\right)f\left(w;\theta^{*}\right)dw.
\]
. Then, $\nabla_{\theta}Q\left(\widehat{\theta}\right)=-\frac{1}{N}\sum_{i=1}^{N}\nabla_{\theta}\ln f\left(w_{i};\widehat{\theta}\right)$ can be decomposed into three terms:

\begin{eqnarray*}
\nabla_{\theta}Q\left(\widehat{\theta}\right) & = & \frac{1}{N}\sum_{i=1}^{N}\left(\nabla_{\theta}\ln f\left(w_{i};\theta^{*}\right)-\nabla_{\theta}\ln f\left(w_{i};\widehat{\theta}\right)\right)+\\
 &  & \int\left(\nabla_{\theta}\ln f\left(w;\theta^{*}\right)\right)f^{*}\left(w\right)dw-\frac{1}{N}\sum_{i}\nabla_{\theta}\ln f\left(w_{i};\theta^{*}\right)+\\
 &  & \int\left(\nabla_{\theta}\ln f\left(w;\theta^{*}\right)\right)\left(f\left(w;\theta^{*}\right)-f^{*}\left(w\right)\right)dw.
\end{eqnarray*}

Concerning the first term, $\nabla_{\theta}\ln f\left(w_{i};\widehat{\theta}\right)-\nabla_{\theta}\ln f\left(w_{i};\theta^{*}\right)=\left(\nabla_{\theta\theta^{\prime}}\ln f\left(w_{i};\exists\overline{\theta}\right)\right)\left(\widehat{\theta}-\theta^{*}\right)$ holds by the mean value theorem. Then, by $\widehat{\theta}-\theta^{*}=O_{p}\left(N^{-\frac{1}{2}}\right)$ and the boundedness of $\nabla_{\theta\theta^{\prime}}\ln f\left(w;\theta\right)$, the first term is $O_{p}\left(N^{-\frac{1}{2}}\right)$. The second term is also $O_{p}\left(N^{-\frac{1}{2}}\right)$ by the central limit theorem and the assumption of finite $Var\left(\nabla_{\theta}\ln f\left(w;\theta^{*}\right)\right)$. The third term is 0 if $f\left(w;\theta^{*}\right)=f^{*}\left(w\right)$, which is satisfied when the distribution $f$ is correctly specified. Consequently, $\nabla_{\theta}Q\left(\widehat{\theta}\right)=O_{p}\left(N^{-\frac{1}{2}}\right)$ holds if the distribution $f$ is correctly specified.

Note that \cite{aitchison1958maximum} originally considered a setting where the objective is $\sum_{i=1}^{N}q_{i}\left(\theta\right)$ (i.e., not normalized by $N$), and showed that the Lagrange multiplier $\widetilde{\lambda}$ satisfies $\widetilde{\lambda}=O_{p}\left(N^{\frac{1}{2}}\right)$. In contrast, we consider the objective $\frac{1}{N}\sum_{i=1}^{N}q_{i}\left(\theta\right)$, and $\widehat{\lambda}=O_{p}\left(N^{-\frac{1}{2}}\right)$ holds.

\subsubsection{Least Squares (LS)\label{subsec:LS}}

\subsubsection*{Intuition}

Concerning the LS problem, it is well known that the problem is essentially equivalent to the MLE if the residuals in the LS follow the normal distribution. Consequently, the values of the Lagrange multipliers should be equal to 0 in the infinite-sample limit when the residuals actually follow a normal distribution.

\subsubsection*{Formal discussion\protect\footnote{When the parameters are identified without introducing the constraint (i.e., $\theta^{*}=\min_{\theta}Q(\theta)$ when $N$ is infinite), $\widehat{\lambda}=O_{p}\left(N^{-\frac{1}{2}}\right)$ holds, even when the residuals do not follow the normal distribution. Previous studies utilized the property for statistical hypothetical testings (cf. \cite{gourieroux1995statistics}, for instance). }}

Here, let $f^{*}\left(r\right)$ be the true distribution of the residuals $\left\{ r\left(w,\theta^{*}\right)\right\} _{w\sim\mathcal{P}}$, where $w\sim\mathcal{P}$ implies that the observation $w$ is drawn from $\mathcal{P}$. Let $\sigma$ be the standard deviation of $\left\{ r\left(w,\theta^{*}\right)\right\} _{w\sim\mathcal{P}}$. We impose the assumption of finite $Var\left(\nabla_{\theta}\left(\ln\phi_{\sigma}\left(r\left(w,\theta^{*}\right)\right)\right)\right)$ and the boundedness of $\nabla_{\theta\theta^{\prime}}\ln\phi_{\sigma}\left(r\left(w,\theta\right)\right)$. First, $Q\left(\theta\right)$ can be reformulated as:

\begin{eqnarray*}
Q\left(\theta\right) & = & \frac{1}{2N}\sum_{i=1}^{N}\left(r\left(w_{i},\theta\right)\right)^{2}\\
 & = & -2\sigma^{2}\frac{1}{2N}\sum_{i=1}^{N}\ln\left(\frac{1}{\sqrt{2\pi}\sigma}\exp\left(-\frac{\left(r\left(w_{i},\theta\right)\right)^{2}}{2\sigma^{2}}\right)\right)+\sigma^{2}\ln\left(\frac{1}{\sqrt{2\pi}\sigma}\right)
\end{eqnarray*}

Let $\phi_{\sigma}\left(r\right)\equiv\frac{1}{\sqrt{2\pi}\sigma}\exp\left(-\frac{r^{2}}{2\sigma^{2}}\right)$, which corresponds to the density function of the normal distribution with standard deviation $\sigma$. We also define $h_{\sigma}\left(w;\theta\right)\equiv\phi_{\sigma}\left(r\left(w;\theta\right)\right)$. By $\int_{-\infty}^{\infty}\frac{1}{\sqrt{2\pi}\sigma}\exp\left(-\frac{r^{2}}{2\sigma^{2}}\right)dr=1$, 
\[
1=\int_{-\infty}^{\infty}\frac{1}{\sqrt{2\pi}\sigma}\exp\left(-\frac{\left(y-\psi\left(x,\theta\right)\right)^{2}}{2\sigma^{2}}\right)dy=\int h_{\sigma}\left(w\equiv\left(x,y\right);\theta\right)dy
\]
 holds for all $x$ and $\theta$. Then, by differentiating the both sides with respect to $\theta$ at $\theta=\theta^{*}$, 
\[
\int\left(\nabla_{\theta}\ln h_{\sigma}\left(w;\theta^{*}\right)\right)h_{\sigma}\left(w;\theta^{*}\right)dy=0
\]
 holds for all $x$, which implies 
\[
\int\left(\nabla_{\theta}\ln h_{\sigma}\left(w;\theta^{*}\right)\right)h_{\sigma}\left(w;\theta^{*}\right)dw=\int\left[\int\left(\nabla_{\theta}\ln h_{\sigma}\left(w;\theta^{*}\right)\right)h_{\sigma}\left(w;\theta^{*}\right)dy\right]dx=0.
\]

Then, $\nabla_{\theta}Q\left(\widehat{\theta}\right)=\nabla_{\theta}\left[-\sigma^{2}\frac{1}{N}\sum_{i=1}^{N}\ln h_{\sigma}\left(w_{i};\widehat{\theta}\right)\right.$$\left.+\sigma^{2}\ln\left(\frac{1}{\sqrt{2\pi}\sigma}\right)\right]$ can be decomposed into three terms:

\begin{eqnarray*}
\nabla_{\theta}Q\left(\widehat{\theta}\right) & = & \sigma^{2}\left[\frac{1}{N}\sum_{i}\left(\nabla_{\theta}\ln h_{\sigma}\left(w_{i};\theta^{*}\right)-\nabla_{\theta}\ln h_{\sigma}\left(w_{i};\widehat{\theta}\right)\right)\right]+\\
 &  & \sigma^{2}\left[\int\left(\nabla_{\theta}\ln h_{\sigma}\left(w;\theta^{*}\right)\right)f^{*}\left(w\right)dw-\frac{1}{N}\sum_{i}\nabla_{\theta}\left(\ln h_{\sigma}\left(w_{i};\theta^{*}\right)\right)\right]+\\
 &  & \sigma^{2}\left[\int\left(\nabla_{\theta}\ln h_{\sigma}\left(w;\theta^{*}\right)\right)\left(h_{\sigma}\left(w;\theta^{*}\right)-f^{*}\left(w\right)\right)dw\right].
\end{eqnarray*}

Concerning the first term, $\nabla_{\theta}\ln\left(h_{\sigma}\left(w_{i};\widehat{\theta}\right)\right)-\nabla_{\theta}\ln\left(h_{\sigma}\left(w_{i};\theta^{*}\right)\right)=\left(\nabla_{\theta\theta^{\prime}}\ln\left(h_{\sigma}\left(w_{i};\exists\overline{\theta}\right)\right)\right)\left(\widehat{\theta}-\theta^{*}\right)$ holds by the mean value theorem. Then, by $\widehat{\theta}-\theta^{*}=O_{p}\left(N^{-\frac{1}{2}}\right)$ and the boundedness of $\nabla_{\theta\theta^{\prime}}\ln f\left(w_{i};\exists\overline{\theta}\right)$, the first term is $O_{p}\left(N^{-\frac{1}{2}}\right)$. The second term is also $O_{p}\left(N^{-\frac{1}{2}}\right)$ by the central limit theorem. The third term is 0 if the residual actually follows the normal distribution, i.e., $f^{*}\left(w\right)=h_{\sigma}\left(w;\theta^{*}\right)$. Consequently, $\nabla_{\theta}Q\left(\widehat{\theta}\right)=O_{p}\left(N^{-\frac{1}{2}}\right)$ holds if the residuals follow a normal distribution.

\subsection{Application in Deep learning}

The discussions so far, which establish $\widehat{\lambda}=O_{p}\left(N^{-\frac{1}{2}}\right)$, were based on a classical statistical setting in which the number of free parameters is smaller than the number of observations $N$. However, in deep learning applications, the number of parameters can be in the millions or even billions, and therefore a direct application of the previous arguments may not be appropriate. Nevertheless, we can still show that $\widehat{\lambda}\approx0$ holds by slightly modifying the earlier reasoning. In this subsection, we focus on the LS problem, since PINNs primarily rely on the LS formulation. Note that analogous arguments also apply to the MLE setting.

First, we choose $\theta^{*}\in\arg\min_{\theta}E_{w}\left[\left(y-\psi\left(x;\theta\right)\right)^{2}\right]\ s.t.\ g_{E}\left(\theta\right)=0,g_{I}\left(\theta\right)\leq0$. Note that the minimizer may not be unique, and we select the one closest to $\widehat{\theta}$. Concerning the three terms discussed in Section \ref{subsec:LS}, the second term becomes close to 0 by the law of large numbers when $N$ is large, and the third term becomes close to zero when the true error distribution is close to normal. Importantly, these arguments continue to hold even when the number of parameters is very large. Concerning the first term, the mean value theorem implies

$\nabla_{\theta}\ln\left(h_{\sigma}\left(w_{i};\widehat{\theta}\right)\right)-\nabla_{\theta}\ln\left(h_{\sigma}\left(w_{i};\theta^{*}\right)\right)=\left(\nabla_{\theta\theta^{\prime}}\ln\left(h_{\sigma}\left(w_{i};\exists\overline{\theta}\right)\right)\right)\left(\widehat{\theta}-\theta^{*}\right)$, where $\overline{\theta}$ lies between $\widehat{\theta}$ and $\theta^{*}$. This suggests that the first term becomes small when if the generalization performance is good and $\widehat{\theta}\approx\theta^{*}$ holds. Consequently, under the conditions that the error distribution is close to normal, the sample size is large, and the generalization performance is good, the values of $\nabla_{\theta}Q\left(\widehat{\theta}\right)$ and $\widehat{\lambda}$ will be close to zero.

Regarding this point, \cite{lu2021physics}, who proposed a deep learning-based method (PINNs with hard constraints) for solving topology optimization problems, reported a numerical example in which the values of $\widehat{\lambda}$ concentrate near zero in a holography problem in optics (Figure 6 of \cite{lu2021physics}). The constrained optimization problem they consider is essentially similar to the LS problem (see also Section \ref{subsec:Mean-Squared-Residual} of the current paper). Although they did not discuss any mechanisms behind the numerical results, our statistical arguments offer a natural interpretation: the Lagrange multipliers tend to be close to zero when the distribution of the residuals is close to normal.

\subsection{Mean Squared Residual\label{subsec:Mean-Squared-Residual}}

In applications of PDE (Partial Differential Equation)-constrained optimizations for solving inverse problems, the problems are sometimes in the following form, which involves an integral:

\begin{eqnarray*}
 & \min_{\theta} & Q\left(\theta\right)=\frac{1}{\left|Area\left(\Omega\right)\right|}\int_{\Omega}\left(y\left(x\right)-\psi\left(x;\theta\right)\right)^{2}dx\\
 & s.t. & g_{E}\left(\theta\right)=0\\
 &  & g_{I}\left(\theta\right)\leq0
\end{eqnarray*}
where $\Omega$ denotes a region and $Area\left(\Omega\right)$ represents its area. For instance, \cite{lu2021physics} consider a holography problem in physics, that is formulated as a constrained optimization problem of this type. Note that $y\left(x\right)-\psi\left(x;\theta\right)$ corresponds to the difference between a target specified by an analyst and a model-predicted value in the problem. The discussions in Section \ref{subsec:LS} imply that the Lagrange multipliers for this problem become equal to zero when the residuals $\left\{ y\left(x\right)-\psi\left(x;\theta\right)\right\} _{x\in\Omega}$ follows a normal distribution.

\section{Implications\label{sec:Implications}}

\subsection{Augmented Lagrange Multiplier (ALM) Method\label{subsec:ALM}}

To simplify the discussion, we consider the Augmented Lagrange Multiplier (ALM) method for an equality-constrained problem of the form $\min_{\theta}Q\left(\theta\right)\ s.t.\ g\left(\theta\right)=0$. Let $L_{\rho}(\theta,\lambda)\equiv Q(\theta)+\lambda^{T}g(\theta)+\frac{\rho}{2}\left\Vert g(\theta)\right\Vert ^{2}$ be the augmented Lagrangian function, where $\rho$ is a penalty parameter. In the ALM, the following steps are iterated until convergence given the initial Lagrange multipliers $\lambda_{0}$ $(k=0,1,2,\cdots)$: $\theta_{k+1}=\min_{\theta}L_{\rho}(\theta,\lambda_{k})$, $\lambda_{k+1}=\lambda_{k}+\rho g_{E}(\theta_{k+1})$. As shown in \cite{bertsekas1982constrained}, there exists a $\overline{\rho_{N}}>0$ such that $\nabla_{\theta}L_{\overline{\rho_{N}}}\left(\widehat{\theta},\widehat{\lambda}\right)>0$ and $\left\Vert \lambda_{k+1}-\widehat{\lambda}\right\Vert \leq\frac{\exists M_{N}}{\rho}\left\Vert \lambda_{k}-\widehat{\lambda}\right\Vert $, i.e., $\left\Vert \lambda_{k}-\widehat{\lambda}\right\Vert \leq\left(\frac{M_{N}}{\rho}\right)^{k}\left\Vert \lambda_{0}-\widehat{\lambda}\right\Vert $ for $\rho\geq\overline{\rho_{N}}$, and choosing $\lambda_{0}$ close to $\widehat{\lambda}$ is desirable to reduce the number of iterations. Note that $\overline{\rho_{N}}$ and $M_{N}$ do not diverge even when $N\rightarrow\infty$, as long as the solution $\min_{\theta}Q\left(\theta\right)=E_{w}\left[q\left(w;\theta\right)\right]\ s.t.\ g\left(\theta\right)=0$ exists.

\cite{birgin2014practical} have recommended the use of 0 as the initial Lagrange multipliers unless we have prior knowledge on the true value of $\widehat{\lambda}$, although formal justifications were not given. Previous studies utilizing ALM-type algorithms set the initial Lagrange multipliers to 0 in practice (e.g., \cite{dener2020training,lu2021physics,basir2023adaptive} for physics constrained neural networks, \cite{sangalli2021constrained} for solving class-imbalanced binary classification). The discussions in Section \ref{sec:Problem-setting} suggest that the Lagrange multipliers of the problem (\ref{eq:constrained_opt_problem}) are close to 0 under the conditions discussed before. Therefore, choosing $\lambda_{0}=0$ is not only a practical heuristic but also statistically justified.

\subsection{Sequential Quadratic Programming and Interior Point Method\label{subsec:SQP_IP} }

Sequential Quadratic Programming (SQP) and Interior Point (IP) methods require specifying not only the initial value of $\theta$ but also the initial Lagrange multipliers. Although previous studies have established superlinear local convergence and global convergence properties for these methods, their practical performance can deteriorate when the initial values are far from the true solution. In practice, the initial multipliers $\lambda_{0}=0$ are typically set to zero or chosen as the least-squares solution to the dual infeasibility,\footnote{When only the equality constraints $g(\theta)=0$ exists, $\lambda_{0}$ is chosen as the solution of a least-square problem $\min_{\theta}\left\Vert \nabla_{\theta}Q\left(\theta_{0}\right)+\lambda^{T}\left(g\left(\theta_{0}\right)\right)\right\Vert _{2}^{2}$, where $\theta_{0}$ denotes the initial values of $\theta$ specified by an analyst.} often combined with safeguard strategies in certain solvers (e.g., IPOPT; \cite{wachter2006implementation}), unless prior knowledge about the true multipliers is available.

The results presented in this study suggest that, under the statistical conditions discussed earlier, the true Lagrange multipliers tend to be close to zero. This implies that choosing $\lambda_{0}=0$ may reduce the number of iterations and function evaluations required for convergence. Please note that, unlike the ALM, the initial choice of $\theta$, is also important for the convergence. Nevertheless, initial $\lambda$ far from the true value may make the convergence slower and more unstable. Numerical results in Section \ref{sec:Numerical-Experiments} support this observation. 

\subsection{Unconstrained optimization with a penalty term (Soft constraint)}

To simplify the discussion, we consider the case in which only equality constraints $g\left(\theta\right)=0$ exist. In some applications, sometimes the following problem is solved instead of the original constrained optimization problem:

\begin{eqnarray}
\min_{\theta} & Q\left(\theta\right)+\rho\left\Vert g\left(\theta\right)\right\Vert _{2}^{2}\label{eq:soft_constraint_opt}
\end{eqnarray}
$\rho\left\Vert g\left(\theta\right)\right\Vert _{2}^{2}$ represents a penalty term that becomes large when $g\left(\theta\right)$ deviates from zero. Let $\widetilde{\theta}$ denote the solution to the problem (\ref{eq:soft_constraint_opt}). 

First, $\widehat{\theta}$ is a local minimizer of $\min_{\theta}Q\left(\theta\right)+\widehat{\lambda}g\left(\theta\right)+\rho\left\Vert g\left(\theta\right)\right\Vert ^{2}$ when $\rho$ is large enough (cf. Theorem 17.5 of \cite{nocedal2006numerical}). Here, $\widehat{\lambda}$ can be regarded as the consistent estimator of $\lambda^{*}=0$, because $\widehat{\lambda}=O_{p}\left(N^{-\frac{1}{2}}\right)$. Thus, by treating the Lagrange multipliers $\lambda$ as nuisance parameters and applying Property 24.8 of \cite{gourieroux1995statistics},\footnote{The property claims that the extremum estimator $\widehat{\beta}$, which is the solution to $\min_{\beta}\widetilde{Q}$$\left(\widehat{\alpha},\beta\right)$, is a consistent estimator of true $\beta$ if $\widehat{\alpha}$ is a consistent estimator of the true nuisance parameter $\alpha$.} $\widetilde{\theta}$ can be viewed as a consistent estimator of $\theta^{*}$, which is the solution of $\min_{\theta}Q\left(\theta\right)=E_{w}\left[q\left(w;\theta\right)\right]\ s.t.\ g\left(\theta\right)=0$.

In deep learning applications, however, such asymptotic arguments may not hold because the number of parameters can be much larger than the number of observations. Even so, we can generally show that the difference between $\widehat{\theta}$ and $\widetilde{\theta}$ remains small under moderate conditions. 

The first\nobreakdash-order optimality conditions for $\widehat{\theta}$ and $\widetilde{\theta}$ imply:

\begin{eqnarray}
\nabla_{\theta}Q\left(\widehat{\theta}\right)+\widehat{\lambda}\left(\nabla_{\theta}g\left(\widehat{\theta}\right)\right) & = & 0\label{eq:FOC_theta1}\\
\nabla_{\theta}Q\left(\widetilde{\theta}\right)+2\rho\left(\nabla_{\theta}g\left(\widetilde{\theta}\right)\right)^{T}\left(g\left(\widetilde{\theta}\right)\right) & = & 0\label{eq:FOC_theta2}
\end{eqnarray}

By Taylor's theorem, $\nabla_{\theta}Q\left(\widetilde{\theta}\right)=\nabla_{\theta}Q\left(\widehat{\theta}\right)+\left(\nabla_{\theta\theta^{\prime}}Q\left(\overline{\theta}\right)\right)\left(\widetilde{\theta}-\widehat{\theta}\right)$ and $g\left(\widetilde{\theta}\right)=g\left(\widehat{\theta}\right)+\left(\nabla_{\theta}g\left(\breve{\theta}\right)\right)\left(\widetilde{\theta}-\widehat{\theta}\right)$ holds, where $\overline{\theta}$ and $\breve{\theta}$ lie between $\widehat{\theta}$ and $\widetilde{\theta}$. Then, (\ref{eq:FOC_theta1}), (\ref{eq:FOC_theta2}), and $g\left(\widehat{\theta}\right)=0$ imply $\left(\nabla_{\theta\theta^{\prime}}Q\left(\overline{\theta}\right)+2\rho\left(\nabla_{\theta}g\left(\widetilde{\theta}\right)\right)^{T}\left(\nabla_{\theta}g\left(\breve{\theta}\right)\right)\right)\left(\widetilde{\theta}-\widehat{\theta}\right)=\widehat{\lambda}\left(\nabla g\left(\widehat{\theta}\right)\right)$. Assuming the nonsingularity of the matrix $\nabla_{\theta\theta^{\prime}}Q\left(\overline{\theta}\right)+2\rho\left(\nabla_{\theta}g\left(\widetilde{\theta}\right)\right)^{T}\left(\nabla_{\theta}g\left(\breve{\theta}\right)\right)$, we obtain $\widetilde{\theta}-\widehat{\theta}=\left(\nabla_{\theta\theta^{\prime}}Q\left(\overline{\theta}\right)+2\rho\left(\nabla_{\theta}g\left(\widetilde{\theta}\right)\right)^{T}\left(\nabla_{\theta}g\left(\breve{\theta}\right)\right)\right)^{-1}\widehat{\lambda}\left(\nabla g\left(\widehat{\theta}\right)\right)$.

The formula suggests that $\widetilde{\theta}-\widehat{\theta}$ is close to 0 when $\widehat{\lambda}$ is close to 0, unless $\nabla_{\theta\theta^{\prime}}Q\left(\overline{\theta}\right)+2\rho\left(\nabla_{\theta}g\left(\widetilde{\theta}\right)\right)^{T}\left(\nabla_{\theta}g\left(\breve{\theta}\right)\right)$ becomes nonsingular. Hence, small $\left\Vert \widehat{\lambda}\right\Vert $ implies small difference between $\widehat{\theta}$ and $\widetilde{\theta}$. 

Since $\widehat{\lambda}$ tends to be close to zero in large samples for MLE and LS under the conditions discussed earlier, the difference between the exact constrained solution and the solution to the penalized unconstrained problem can be very small. This provides a theoretical explanation for why penalty\nobreakdash-based approaches often perform well in practice, even when the penalty parameter $\rho$ is not extremely large.

Although the difference between the two solutions vanishes as $\rho\rightarrow\infty$, it is well known that excessively large values of $\rho$ can lead to numerical instability, making reliable optimization difficult. The results of the present study suggest that penalty\nobreakdash-based methods can still perform well with moderate values of $\rho$ when the true Lagrange multipliers $\widehat{\lambda}$ are small.

\medskip{}

\section{Numerical Experiments\label{sec:Numerical-Experiments}}

This section presents numerical experiments designed to demonstrate that choosing $\lambda_{0}=0$ is effective in the statistical applications discussed in this paper. We consider both the Augmented Lagrange Multiplier (ALM) algorithm and the Interior Point (IP) algorithm. All experiments are implemented in Julia. For the IP method, we use the IPOPT package (v1.13.0) in Julia 1.12.3. Additional details of the experimental setup are provided in Appendix \ref{subsec:Details-numerical-experiments}.

\subsection{Constrained regression problems}

We first conduct numerical experiments focusing on constrained regression problems---specifically, linear regression and logistic regression---as in \cite{na2025statistical}. These correspond to LS and MLE problems, respectively. In both cases, we impose a nonlinear constraint $\left\Vert \theta\right\Vert _{2}^{2}=b$, which is consistent with the data\nobreakdash-generating process.

Table \ref{tab:Results-regression-ALM} reports the results for the ALM. For each problem setting, we vary the initial Lagrange multiplier $\lambda_{0}$ and examine its effect on the number of main iterations and the number of objective evaluations. The results clearly indicate that choosing $\lambda_{0}=0$ minimizes both the number of main iterations and the number of objective evaluations.

\begin{table}[H]
\caption{Results of numerical experiments (Constrained regressions; Augmented Lagrangian Multiplier (ALM) Method)\label{tab:Results-regression-ALM}}

\begin{centering}
\begin{tabular}{ccccccc}
\hline 
\multirow{2}{*}{Obj} & \multirow{2}{*}{$\lambda_{0}$} & \multicolumn{2}{c}{\# main iter} & \multicolumn{2}{c}{\# obj eval} & \multirow{2}{*}{$\widehat{\lambda}$}\tabularnewline
 &  & Mean & Std & Mean & Std & \tabularnewline
\hline 
\multirow{5}{*}{Lin} & 0 & 5.3 & 0.483 & 120.2 & 8.244 & \multirow{5}{*}{-0.004}\tabularnewline
 & 1 & 7 & 0 & 177.5 & 6.468 & \tabularnewline
 & 10 & 8.4 & 0.516 & 216.8 & 5.007 & \tabularnewline
 & -1 & 7 & 0 & 177.5 & 4.378 & \tabularnewline
 & -10 & 8 & 0 & 236.6 & 8.618 & \tabularnewline
\hline 
\multirow{5}{*}{Logit} & 0 & 2 & 0 & 109.4 & 6.186 & \multirow{5}{*}{-0.001}\tabularnewline
 & 1 & 3 & 0 & 145.8 & 5.712 & \tabularnewline
 & 10 & 4 & 0 & 168.1 & 12.53 & \tabularnewline
 & -1 & 3 & 0 & 162.7 & 11.87 & \tabularnewline
 & -10 & 3.7 & 0.483 & 237.2 & 14.831 & \tabularnewline
\hline 
\end{tabular}
\par\end{centering}
{\footnotesize Notes. Obj=``Lin'' and ``Logit'' correspond to the linear regression and the logistic regression, respectively. Based on 10 trials using random initial values of $\theta$ and randomly generated datasets given each choice of $\lambda_{0}$.}{\footnotesize\par}
\end{table}

Table \ref{tab:Results-regression-IP} presents the results for the IP method. We again observe the tendency that $\lambda_{0}=0$ outperforms other choices. Additional numerical results are provided in Appendix \ref{subsec:Additional-results}, and they exhibit similar patterns.

\begin{table}[H]
\caption{Results of numerical experiments (Constrained regressions; Interior Point (IP) method)\label{tab:Results-regression-IP}}

\begin{centering}
\begin{tabular}{ccccccc}
\hline 
\multirow{2}{*}{Obj} & \multirow{2}{*}{$\lambda_{0}$} & \multicolumn{2}{c}{\# obj eval} & \multicolumn{2}{c}{\# Hessian eval} & \multirow{2}{*}{$\widehat{\lambda}$}\tabularnewline
 &  & Mean & Std & Mean & Std & \tabularnewline
\hline 
\multirow{5}{*}{Lin} & 0 & 11.4 & 2.757 & 8.3 & 1.059 & \multirow{5}{*}{-0.004}\tabularnewline
 & 1 & 12.4 & 4.742 & 8.2 & 1.317 & \tabularnewline
 & 10 & 13 & 3.528 & 9.6 & 0.699 & \tabularnewline
 & -1 & 13 & 3.83 & 8.3 & 1.337 & \tabularnewline
 & -10 & 14 & 6.549 & 10 & 1.7 & \tabularnewline
\hline 
\multirow{5}{*}{Logit} & 0 & 9.1 & 0.568 & 8.1 & 0.568 & \multirow{5}{*}{-0.001}\tabularnewline
 & 1 & 17.2 & 1.814 & 14.1 & 1.101 & \tabularnewline
 & 10 & 17.8 & 3.293 & 14.1 & 1.197 & \tabularnewline
 & -1 & 12.9 & 4.841 & 9.8 & 1.229 & \tabularnewline
 & -10 & 13.1 & 2.514 & 11.1 & 1.449 & \tabularnewline
\hline 
\end{tabular}
\par\end{centering}
{\footnotesize Notes. Obj=``Lin'' and ``Logit'' correspond to the linear regression and the logistic regression,respectively.}{\footnotesize\par}

{\footnotesize Based on 10 trials using random initial values of $\theta$ and randomly generated datasets given each choice of $\lambda_{0}$.}{\footnotesize\par}
\end{table}

\subsection{Dynamic discrete choice model estimation}

We next conduct numerical experiments by focusing on dynamic discrete choice (DDC) model estimation, using the bus engine replacement data from \cite{rust1987optimal}, which has been treated as a benchmark model and dataset in the economics literature (cf. \cite{su2012constrained,iskhakov2016comment}). Note that estimations of dynamic discrete choice models is closely related to inverse reinforcement learning, which has been extensively considered in recent machine learning studies, as discussed \cite{sanghvi2021inverse}. In this problem, we estimate the utility parameters of a forward\nobreakdash-looking agent who owns buses and decides when to replace their engines. The objective function corresponds to a likelihood function, and the constraints correspond to equalities derived from the agent’s Bellman equation. As discussed in Appendix \ref{subsec:Details-numerical-experiments}, we also impose several inequality constraints to exclude unrealistic parameter domains.

Table \ref{tab:Results-DDC} shows the results when we apply the IP. As in the case of the constrained regression problems, the choice of $\lambda_{0}=0$ usually outperforms other values of $\lambda_{0}$. Note that the table also reports the results obtained when we employ the default choice of $\lambda$. The numerical results suggest that the simple choice of $\lambda_{0}=0$ can outperform the default setting, which incorporates several safeguard strategies.

\begin{table}[H]
\caption{Results of numerical experiments (Dynamic discrete choice model estimation; Interior Point (IP) method)\label{tab:Results-DDC}}

\begin{centering}
\begin{tabular}{cccccccc}
\hline 
\multirow{2}{*}{$N$} & \multirow{2}{*}{$\lambda_{0}$} & \multicolumn{2}{c}{\# obj eval} & \multicolumn{2}{c}{\# Hessian eval} & \multicolumn{2}{c}{$\widehat{\lambda}$}\tabularnewline
 &  & Mean & Std & Mean & Std & Mean & Std\tabularnewline
\hline 
 & Default & 22.2 & 6.233 & 17.7 & 2.83 & \multirow{6}{*}{-0.027} & \multirow{6}{*}{0.164}\tabularnewline
 & 0 & 21.7 & 3.592 & 17.8 & 1.814 &  & \tabularnewline
All & 1 & 19.4 & 2.366 & 16.4 & 1.776 &  & \tabularnewline
(8260) & 10 & 17.6 & 3.307 & 15.3 & 1.947 &  & \tabularnewline
 & -1 & 22.7 & 9.9 & 18.1 & 3.957 &  & \tabularnewline
 & -10 & 34.2 & 17.85 & 24 & 8.563 &  & \tabularnewline
\hline 
\multirow{6}{*}{1000} & Default & 32.5 & 16.555 & 22.3 & 5.87 & \multirow{6}{*}{-0.004} & \multirow{6}{*}{0.02}\tabularnewline
 & 0 & 22.2 & 5.884 & 18.4 & 3.204 &  & \tabularnewline
 & 1 & 23.3 & 8.807 & 19.2 & 6.215 &  & \tabularnewline
 & 10 & 24.6 & 5.461 & 19.6 & 3.534 &  & \tabularnewline
 & -1 & 28.1 & 8.103 & 21.5 & 4.79 &  & \tabularnewline
 & -10 & 44.4 & 13.882 & 29.9 & 6.226 &  & \tabularnewline
\hline 
\multirow{6}{*}{500} & Default & 32 & 12.156 & 23 & 7.483 & \multirow{6}{*}{-0.002} & \multirow{6}{*}{0.009}\tabularnewline
 & 0 & 26.4 & 8.058 & 20.3 & 4.423 &  & \tabularnewline
 & 1 & 28 & 12.019 & 21 & 7.986 &  & \tabularnewline
 & 10 & 34.4 & 10.977 & 22.8 & 6.374 &  & \tabularnewline
 & -1 & 37.4 & 17.43 & 26.1 & 7.767 &  & \tabularnewline
 & -10 & 46.2 & 11.448 & 32.2 & 7.997 &  & \tabularnewline
\hline 
\end{tabular}
\par\end{centering}
{\footnotesize Notes. Based on 10 trials using random initial values of $\lambda_{0}$ and randomly generated datasets.}{\footnotesize\par}

{\footnotesize$N=8260$ is the case where we use all the samples.}{\footnotesize\par}

{\footnotesize ``Default'' represents the case using the default choice of $\lambda_{0}$ in the IPOPT package.}{\footnotesize\par}
\end{table}

\section{Conclusions\label{sec:Conclusions}}

The current study has investigated the property of Lagrange multipliers in constrained Maximum Likelihood Estimation (MLE) and Least Squares (LS) problems---which frequently appear in various applications---from the perspective of numerical optimization. Building on large-sample theory in statistics, we can utilize a property that Lagrange multiplier take values close to zero under some conditions. The property would be useful for further investigation of more efficient numerical optimization methods, and we leave this direction for future research.

\section*{Declarations}

\paragraph*{Funding}

This study was supported by JSPS KAKENHI Grant Number JP24K22629.

\appendix

\section{Appendix\label{sec:Appendix}}

\subsection{Details of the numerical experiments\label{subsec:Details-numerical-experiments}}

The replication code is available at \url{https://github.com/takeshi-fukasawa/Constrained_opt_MLE_LS}.

\subsubsection{Constrained regression problems}

\paragraph*{Linear regression}

$\xi_{b}$ is generated by $\xi_{b}=\xi_{a}\theta^{*}+\epsilon$, where $\epsilon\sim N(0,1)$. $q(w,\theta)$ is in the form of $q(w,\theta)=\frac{1}{2}\left(\xi_{a}^{T}\theta-\xi_{b}\right)^{2}$.

\paragraph*{Logistic regression}

$\xi_{b}\in\{-1,1\}$ is generated by a probability $P\left(\xi_{b}|\xi_{a}\right)=\frac{\exp\left(\xi_{b}\cdot\xi_{a}\theta^{*}\right)}{1+\exp\left(\xi_{b}\cdot\xi_{a}\theta^{*}\right)}$. $q(w,\theta)$ is in the form of $q(w,\theta)=\log\left(1+\exp\left(-\xi_{b}\cdot\xi_{a}^{T}\theta\right)\right)$.

In both types of problems, $\xi_{a}$ is drawn from $N\left(0,5I+\Sigma_{a}\right)$, where (i)$\Sigma_{a}=I$ (Identity), (ii)$\left[\Sigma_{a}\right]_{i,j}=r^{|i-j|}=0.5^{|i-j|}$ (Toeplitz), (iii) $\left[\Sigma_{a}\right]_{i,j}=r=0.5$ for $i\neq j$ and $\left[\Sigma_{a}\right]_{i,i}=1$ (Equi-correlation).\footnote{Such specifications were also considered in \cite{Chen2020}.} The results in the main part of the current paper corresponds to the first case, and the results under the second and the third settings are shown in Appendix\ref{subsec:Additional-results}. True parameter values $\theta^{*}$, which are $n_{\theta}$-dimensional, is linearly spaced between 0 and 1.

When we impose the nonlinear constraint $\left\Vert \theta\right\Vert _{2}^{2}=b$, we take $b=\left\Vert \theta^{*}\right\Vert _{2}^{2}$. In Appendix \ref{subsec:Additional-results}, we alternatively consider linear constraints $A\theta=b$. The values of $A$'s elements are drawn from the standard normal distribution. Note that $b$ is taken so that $b=A\theta^{*}$. As in \cite{na2025statistical}, we assume that the number of constraints is $\left\lceil \sqrt{n_{\theta}}\right\rceil $. 

We let the number of parameters $n_{\theta}=30$, and the number of data samples $N=1000$. Concerning the initial parameter values specified in the ALM and IP algorithms, they are drawn from $U(-1,1)$.

\subsubsection{Dynamic discrete choice model estimation}

Concerning the dynamic discrete choice model estimation of bus engine replacement model (\cite{rust1987optimal}), we treat the expected value functions$\left\{ EV\left(a,x\right)\right\} _{a\in\{0,1\},x\in\mathcal{X}}$ as variables, and impose the following constraint:

\begin{eqnarray*}
EV\left(a,x\right) & = & \sum_{x^{\prime}}\left(f_{\theta_{f}}\left(x^{\prime}|x,a\right)\right)\log\left(\sum_{a\in\{0,1\}}\exp\left(u_{\theta_{u}}\left(a,x\right)+\beta EV\left(a,x\right)\right)\right)
\end{eqnarray*}
Here, $a\in\{0,1\}$ represents the agent's action: $a=1$ implies the bus engine replacement, and $a=0$ implies no replacement. The state variable $x\in\mathcal{X}$ represents the mileage accumulated over time. We assume the set of states $\mathcal{X}$ is discrete, and $x$ can take values in $\left\{ 0,\delta,2\delta,\cdots,\overline{x}_{max}\right\} $, where $\overline{x}_{max}=(n_{max}-1)\delta$. $\beta$ denotes the discount factor, and $\theta_{f}$ and $\theta_{u}$ denote parameters of state transitions and the agent's utility function. Note that the formula is derived from the Bellman equation using the integrated value function $V\left(x\right)$: 

\[
V\left(x\right)=\log\left(\sum_{a\in\{0,1\}}\exp\left(u_{\theta_{u}}\left(a,x\right)+\beta\sum_{x^{\prime}}V\left(x^{\prime}\right)f_{\theta_{f}}\left(x^{\prime}|x,a\right)\right)\right)
\]

The utility function of the agent is given by:

\begin{eqnarray*}
u_{\theta_{u}}\left(a_{t},x_{t}\right) & \equiv & \begin{cases}
0 & \text{if}\ a_{t}=1\\
\theta_{u}^{(0)}-\theta_{u}^{(1)}x_{t} & \text{if}\ a_{t}=0
\end{cases}
\end{eqnarray*}

State transition is given by:

\begin{eqnarray*}
f_{\theta_{f}}\left(x_{t+1}|x_{t},a_{t}\right) & = & \begin{cases}
\theta_{f}^{(2)}\exp\left(\theta_{f}^{(2)}\left(x_{t+1}-x_{t}\right)\right) & \text{if}\ a_{t}=0\text{\ and\ }x_{t+1}\geq x_{t}\\
\theta_{f}^{(2)}\exp\left(\theta_{2}x_{t+1}\right) & \text{if}\ a_{t}=1\text{\ and\ }x_{t+1}\geq0\\
0 & \text{otherwise}
\end{cases}
\end{eqnarray*}

The conditional choice probability is given by:

\begin{eqnarray*}
q\left(w,\theta\right) & = & \frac{\exp\left(u_{\theta_{u}}\left(a_{t},x_{t}\right)+\beta EV\left(a_{t},x_{t}\right)\right)}{\sum_{a_{t}^{\prime}\in\{0,1\}}\exp\left(u_{\theta_{u}}\left(a_{t}^{\prime},x_{t}\right)+\beta EV\left(a_{t}^{\prime},x_{t}\right)\right)}
\end{eqnarray*}
where $w_{t}=\left(x_{t},a_{t}\right)$. 

The current study conducted numerical experiments by modifying the replication code of \cite{rust1987optimal} written by Florian Oswald.\footnote{The code is available at \url{https://github.com/floswald/Zurcher.jl}.} As in the replication code, we additionally impose the following inequality constraints:
\begin{itemize}
\item $\theta_{u}^{(0)},\theta_{u}^{(1)}\geq0$
\item $-50\leq EV\left(x,a\right)\leq50\ \forall\left(x,a\right)$
\end{itemize}
The former inequalities are imposed to choose appropriate parameter values in the estimation. The latter inequality is intended to avoid numerical instability.

Concerning the initial parameter values specified in the ALM and IP algorithms are drawn from $U(-10,10)$.

We assume the values of $\theta_{f}$ and $\beta$ are known, and we estimate parameters $\theta=\left(\theta_{u},\left\{ EV\left(a,x\right)\right\} _{a\in\{0,1\},x\in\mathcal{X}}\right)$ by solving the constrained optimization. Concerning the known parameters, we let $n_{max}=175,\beta=0.999$. Regarding $\delta$ and $\theta_{u}$, we use the default parameter settings in the replication code.

\subsubsection{ALM and IP algorithms}

In the ALM algorithm (cf. Section \ref{subsec:ALM}), we let $\rho=1$. Note that the value of $\rho$ is set to be constant throughout the iteration. We assume the iteration converges if the norm $\left\Vert g\left(\theta\right)\right\Vert _{2}$ is smaller than 1E-6. For the optimization step in each iteration, we use Optim.optimize function (LBFGS) in Julia. Regarding the IP algorithm, we employ the default parameter settings in the IPOPT package in Julia, except for the initial Lagrange multipliers $\lambda_{0}$.

\subsection{Additional results of numerical experiments\label{subsec:Additional-results}}

Tables \ref{tab:Results-regression-linear-constraints}, \ref{tab:Results-regression-alternative-DGP}, and \ref{tab:Results-misspecified} show the additional results of the numerical experiments. Table \ref{tab:Results-regression-linear-constraints} shows the numerical results when we alternatively impose linear constraints $A\theta=b$, rather than the nonlinear constraint $\left\Vert \theta\right\Vert ^{2}=b$. In Table \ref{tab:Results-regression-alternative-DGP}, we consider alternative data generating processes concerning $\xi_{b}$, by changing the values of $\Sigma$ (See Appendix \ref{subsec:Details-numerical-experiments} for details). Finally, Table \ref{tab:Results-misspecified} shows the results when the true $\xi_{b}$ is drawn based on the normal distribution (i.e. $P\left(\xi_{b}=1|\xi_{a}\right)=\int_{-\infty}^{\xi_{a}\theta^{*}}\phi(t)dt$, where $\phi$ denotes the density function of the standard normal distribution), but we use the logistic regression to estimate parameters. The setting corresponds to the case where the distribution is misspecified in the MLE. However, even in this setting, the values of $\lambda$ are relatively small, and $\lambda_{0}=0$ outperforms the other choices of $\lambda_{0}$.

\begin{table}[H]
\caption{Results of numerical experiments (Constrained regressions; Linear constraints $A\theta=b$)\label{tab:Results-regression-linear-constraints}}

\begin{centering}
\begin{tabular}{cccccccccccc}
\hline 
\multirow{3}{*}{Obj} & \multirow{3}{*}{$\lambda_{0}$} & \multicolumn{4}{c}{ALM} & \multicolumn{4}{c}{IP} & \multicolumn{2}{c}{$\widehat{\lambda}$}\tabularnewline
\cline{3-12}
 &  & \multicolumn{2}{c}{\# main iter} & \multicolumn{2}{c}{\# obj eval} & \multicolumn{2}{c}{\# obj eval} & \multicolumn{2}{c}{\# Hessian eval} &  & \tabularnewline
 &  & Mean & Std & Mean & Std & Mean & Std & Mean & Std & Mean & Std\tabularnewline
\hline 
\multirow{5}{*}{Lin} & 0 & 8.2 & 1.549 & 221.8 & 39.321 & 2 & 0 & 1 & 0 & \multirow{5}{*}{0.000} & \multirow{5}{*}{0.014}\tabularnewline
 & 1 & 11 & 2.16 & 320 & 55.584 & 2 & 0 & 1 & 0 &  & \tabularnewline
 & 10 & 12.4 & 2.319 & 411.2 & 56.245 & 2 & 0 & 1 & 0 &  & \tabularnewline
 & -1 & 10.9 & 1.912 & 318.8 & 52.91 & 2 & 0 & 1 & 0 &  & \tabularnewline
 & -10 & 12.4 & 2.319 & 408.2 & 66.092 & 2 & 0 & 1 & 0 &  & \tabularnewline
\hline 
\multirow{5}{*}{Logit} & 0 & 3.1 & 0.316 & 247.4 & 29.003 & 18.3 & 5.908 & 9 & 1.826 & \multirow{5}{*}{-0.001} & \multirow{5}{*}{0.002}\tabularnewline
 & 1 & 4.4 & 0.516 & 417.4 & 43.998 & 18.4 & 6.168 & 9 & 1.826 &  & \tabularnewline
 & 10 & 4.9 & 0.568 & 690.2 & 99.229 & 18.4 & 6.204 & 9 & 1.826 &  & \tabularnewline
 & -1 & 4.6 & 0.516 & 414.9 & 39.357 & 18.7 & 6.55 & 9 & 1.826 &  & \tabularnewline
 & -10 & 4.9 & 0.316 & 724.9 & 108.834 & 18.7 & 6.201 & 9 & 1.826 &  & \tabularnewline
\hline 
\end{tabular}
\par\end{centering}
\raggedright{}{\footnotesize Notes. Obj=``Lin'' and ``Logit'' correspond to the linear regression and the logistic regression, respectively.}{\footnotesize\par}

{\footnotesize Based on 10 trials using random initial values of $\theta$ and randomly generated datasets given each choice of $\lambda_{0}$. When alternatively using the default value of $\lambda_{0}$ in IPOPT package, the current study obtained the same results as in the case of $\lambda_{0}=0$.}{\footnotesize\par}

\end{table}

\begin{table}[H]
\caption{Results of numerical experiments (Constrained regressions;Alternative $\Sigma$)\label{tab:Results-regression-alternative-DGP}}

\begin{centering}
\begin{tabular}{cccccccccccc}
\hline 
\multirow{3}{*}{Sigma} & \multirow{3}{*}{Obj} & \multirow{3}{*}{$\lambda_{0}$} & \multicolumn{4}{c}{ALM} & \multicolumn{4}{c}{IP} & \multicolumn{1}{c}{$\text{\ensuremath{\widehat{\lambda}}}$}\tabularnewline
\cline{4-12}
 &  &  & \multicolumn{2}{c}{\# main iter} & \multicolumn{2}{c}{\# obj eval} & \multicolumn{2}{c}{\# obj eval} & \multicolumn{2}{c}{\# Hessian eval} & \tabularnewline
 &  &  & Mean & Std & Mean & Std & Mean & Std & Mean & Std & \tabularnewline
\hline 
\multirow{5}{*}{Toeplitz} & \multirow{5}{*}{Logit} & 0 & 2.1 & 0.316 & 157.6 & 13.426 & 9.9 & 0.876 & 8.8 & 0.632 & \multirow{5}{*}{-0.001}\tabularnewline
 &  & 1 & 3 & 0 & 209.1 & 13.395 & 17.9 & 4.149 & 14.6 & 1.265 & \tabularnewline
 &  & 10 & 4 & 0 & 222 & 14.43 & 18.6 & 4.789 & 14.2 & 1.033 & \tabularnewline
 &  & -1 & 3 & 0 & 226.3 & 8.731 & 12.7 & 4.218 & 10.3 & 1.337 & \tabularnewline
 &  & -10 & 3.1 & 0.316 & 340 & 34.775 & 12.7 & 2.163 & 11.2 & 1.549 & \tabularnewline
\hline 
\multirow{5}{*}{Equi-corr} & \multirow{5}{*}{Logit} & 0 & 2 & 0 & 153.9 & 16.299 & 10 & 0.471 & 9 & 0.471 & \multirow{5}{*}{-0.001}\tabularnewline
 &  & 1 & 3 & 0 & 196.3 & 13.191 & 16.8 & 2.7 & 14 & 1.247 & \tabularnewline
 &  & 10 & 4 & 0 & 213.4 & 16.854 & 17.5 & 4.673 & 14.1 & 1.197 & \tabularnewline
 &  & -1 & 3 & 0 & 200.2 & 14.428 & 15.6 & 7.183 & 11 & 1.886 & \tabularnewline
 &  & -10 & 3 & 0 & 319.8 & 20.11 & 13.8 & 4.614 & 11.6 & 1.776 & \tabularnewline
\hline 
\end{tabular}
\par\end{centering}
{\footnotesize Notes. Obj=``Lin'' and ``Logit'' correspond to the linear regression and the logistic regression, respectively.}{\footnotesize\par}

{\footnotesize Based on 10 trials using random initial values of $\theta$ and randomly generated datasets given each choice of $\lambda_{0}$. When alternatively using the default value of $\lambda_{0}$ in IPOPT package, the current study obtained the same results as in the case of $\lambda_{0}=0$.}{\footnotesize\par}

\end{table}

\begin{table}[H]
\caption{Results of numerical experiments (Constrained regressions; Case with Misspecified distribution)\label{tab:Results-misspecified}}

\begin{centering}
\begin{tabular}{ccccccccccc}
\hline 
\multirow{3}{*}{Obj} & \multirow{3}{*}{$\lambda_{0}$} & \multicolumn{4}{c}{ALM} & \multicolumn{4}{c}{IP} & \multicolumn{1}{c}{$\widehat{\lambda}$}\tabularnewline
\cline{3-11}
 &  & \multicolumn{2}{c}{\# main iter} & \multicolumn{2}{c}{\# obj eval} & \multicolumn{2}{c}{\# obj eval} & \multicolumn{2}{c}{\# Hessian eval} & \tabularnewline
 &  & Mean & Std & Mean & Std & Mean & Std & Mean & Std & \tabularnewline
\hline 
\multirow{5}{*}{Logit} & 0 & 3 & 0 & 109.3 & 7.602 & 9 & 0.667 & 8 & 0.667 & \multirow{5}{*}{0.004}\tabularnewline
 & 1 & 3 & 0 & 135.2 & 6.426 & 16.7 & 2.111 & 14.1 & 0.994 & \tabularnewline
 & 10 & 4 & 0 & 169.7 & 7.088 & 18.5 & 4.767 & 14.8 & 2.15 & \tabularnewline
 & -1 & 3 & 0 & 155 & 8.206 & 10.8 & 1.398 & 9.3 & 1.059 & \tabularnewline
 & -10 & 4 & 0 & 224.8 & 18.931 & 15.4 & 8.618 & 11.2 & 2.098 & \tabularnewline
\hline 
\end{tabular}
\par\end{centering}
{\footnotesize Notes. Obj=``Lin'' and ``Logit'' correspond to the linear regression and the logistic regression, respectively.}{\footnotesize\par}

{\footnotesize Based on 10 trials using random initial values of $\theta$ and randomly generated datasets given each choice of $\lambda_{0}$. When alternatively using the default value of $\lambda_{0}$ in IPOPT package, the current study obtained the same results as in the case of $\lambda_{0}=0$.}{\footnotesize\par}

\end{table}

\bibliographystyle{plain}
\bibliography{literature_Lagrangian}
\end{document}